# A Universal Critical Density Underlying the Physics of Electrons at the LaAlO$_3$/SrTiO$_3$ Interface

Arjun Joshua, S. Pecker, J. Ruhman, E. Altman and S. Ilani

*Department of Condensed Matter Physics, Weizmann Institute of Science, Rehovot 76100, Israel.*

**The two-dimensional electron system formed[1] at the interface between the insulating oxides LaAlO$_3$ and SrTiO$_3$ exhibits ferromagnetism[2-5], superconductivity[6,7], and a wide range of unique magnetotransport properties[7-13]. A key challenge is to find a unified microscopic mechanism that underlies these emergent phenomena. Here we show that a universal Lifshitz transition between d-orbitals lies at the core of the observed transport phenomena in this system. Our measurements find a critical electronic density at which the transport switches from single to multiple carriers. This density has a universal value, independent of the LaAlO$_3$ thickness and electron mobility. The characteristics of the transition, its universality, and its compatibility with spectroscopic measurements[14,15] establish it as a transition between d-orbitals of different symmetries. A simple band model, allowing for spin-orbit coupling at the atomic level, connects the observed universal transition to a range of reported magnetotransport properties[9-12]. Interestingly, we also find that the maximum of the superconducting transition temperature occurs at the same critical transition, indicating a possible connection between the two phenomena. Our observations demonstrate that orbital degeneracies play an important role in the fascinating behavior observed so far in these oxides.**

A key feature of the electron liquid at the LaAlO$_3$/SrTiO$_3$ (LAO/STO) interface is that its ground state can be tuned using field-effect gating[16]. Although the origin of the carriers is presently under debate[14,17-20], it is still possible to tune their density and thereby explore a rich phase diagram of fundamental phenomena. Superconductivity was shown to emerge at a certain carrier density, concomitantly with the appearance of strong spin-orbit interactions[9]. At higher densities the critical temperature of superconductivity was shown to slowly decrease[7,8] together with a similar decrease in spin-orbit strength[10]. A growing body of theoretical works[21-24] relate the observed phenomena to electrons populating several d-bands at the interface[14,15,25,26]. Indeed, transport[8,11,12], optical[27] and heat conductivity[13] measurements have found that multiple types of carriers exist in this system, and suggested that they originate from carriers that are localized at different distances from the interface.

In this paper we show that there is a universal origin for the various transport phenomena described above: a transition between single and multiple d-orbitals population. We present results from four samples differing in their LAO layer thickness and spanning an order of magnitude in mobility, all showing similar behavior. Three samples were grown at $T = 800°C$ and the last at $T = 650°C$ to yield higher mobility[11]. Transport was measured using standard back-gated Hall bars with widths varying from $100\mu m$ to $500\mu m$ and ac currents of $5-50nA$. Most measurements were done at $T = 4.2K$, apart from those of superconductivity that were done down to $50mK$. The longitudinal (Hall) resistivity was symmetrized (anti-symmetrized) with respect to magnetic field to eliminate small spurious effects from lithographic imperfections.

All samples measured showed a critical gate voltage across which their transport characteristics changed significantly. Figure 1a shows the measured $\rho_{XY}$ as a function of magnetic field, $B$, at various gate voltages for sample 1. Starting from the lowest gate voltage $V_G = -40V$ and up to a critical value $V_C = 40V$, $\rho_{XY}$ is linear in $B$ with a slope that decreases with increasing $V_G$. This behavior is expected from the field-effect tuning of carrier density. However, as we cross $V_C$ the shape of the Hall traces changes considerably; They become kinked, with the kink position progressively moving to lower $B$ with increasing $V_G$. Zooming-in on the low field Hall slopes (Fig. 1b) we find that they too undergo a significant change at $V_C$, from slopes that gradually decrease with increasing $V_G$ below $V_C$, to slopes that are roughly constant with even a slight increase above $V_C$ (inset Fig. 1b). Thus, both the low and high field Hall characteristics signify that a distinct transition occurs in the system at $V_C$.

The critical transition becomes even more striking when we plot the Hall coefficient $R_H = \frac{\rho_{XY}}{B}$ as a function of $B$ and $V_G$ (Fig. 1c). Below $V_C$, $R_H$ is roughly a constant as a function of $B$, whose value decreases with increasing $V_G$. Above $V_C$, however, the curves acquire a qualitatively different form: They first increase with field at low fields, then peak at an intermediate field, $B_P$, and finally drop down at high fields yielding an overall bell-like shape, whose width we define as $B_W$ (Fig. 1c caption). The characteristic field, $B_W$, progressively decreases as $V_G$ is increased. This trend in $R_H$ is echoed by the longitudinal resistance, $\rho_{XX}$ (Fig. 1d). Below $V_C$, $\rho_{XX}$ is practically independent of $B$, but above $V_C$, $\rho_{XX}$ remains flat only up to an intermediate characteristic field after which it rises continuously. Just like $B_W$, this characteristic field becomes smaller with increasing $V_G$.

The self-similarity of the transport traces on each side of the critical point suggests that they can be grouped into two generic forms. Below $V_C$, the curves have almost identical shapes up to an offset (Figs. 2a and 2b). Above $V_C$, the curves (Figs. 1c and 1d) differ substantially in their dependence on $B$. However, if for every gate voltage we add a y-offset, scale the curve height, and plot it as a function of the scaled magnetic field, $B/B_W$, we observe that all $R_H$ curves perfectly collapse onto a single trace (Fig. 2c). This suggests that the Hall resistance above $V_C$ is described by a generic functional form. Interestingly, if we use the field with which we scaled $R_H$ to scale the $\rho_{XX}$ trace at the same gate voltage, we find that all the $\rho_{XX}$ traces also collapse perfectly (Fig. 2d). The fact that $\rho_{XY}$ and $\rho_{XX}$ scale together demonstrates their strong connection.

In Fig. 2e we plot the dependence of $B_W$ on the carrier density, extracted from the high-field Hall slope. This field increases with decreasing density and becomes immeasurably high when we approach the critical point. If we plot its inverse, $1/B_W$, as a function of the density, we see that it extrapolates to zero at $n_C$, the critical density that marks the onset of the observed transition (Fig. 2f).

What is the nature of the observed critical point? Below $V_C$ the behavior is easy to understand: $\rho_{XY}$ is linear with $B$, its slope is inversely proportional to $V_G$, and concomitantly $\rho_{XX}$ hardly changes with $B$. This simple behavior is exactly what one expects from single-carrier transport. Above $V_C$, on the other hand, the transport shows a much richer structure: the Hall traces are S-shaped and $\rho_{XX}$ shows strong positive magnetoresistance, both suggestive of two-carrier transport. Similar S-shaped Hall traces have been observed in several studies [8,10,12,13], typically done at high carrier densities, and were indeed interpreted as transport by two kinds of carriers. This hints that the critical point corresponds to a transition from one to two-carrier transport.

A simple two-carrier transport model (supplementary material A3) can quantitatively explain the observed transport traces above the critical density, as well as the observed correlation between $\rho_{XY}$ and $\rho_{XX}$. Specifically, we can use this model to determine the densities of the two carriers. Assuming that these carriers have significantly different mobilities, the asymptotic value of $R_H$ at zero field gives the density of the high-mobility carriers, $n_{hi} = 1/eR_0$, whereas its value at large fields yields the total density of both carriers, $n_{total} = 1/eR_\infty$, ($e$ is the electron charge). In Fig. 3a we extract these two quantities, $n_{hi}$ and $n_{total}$, from the Hall coefficients measured at $B = 0T$ and $B = 14T$, and plot them against $V_G$. For voltages below the critical point we see that all the carriers are of the high-mobility type, implied by the fact that $n_{total} \approx n_{hi}$, and that their density is continuously tuned by the gate up to a critical value, $n_C$. After crossing the critical point, $n_{hi}$ saturates at $n_C$, whereas $n_{total}$ continues to rise. This means that only lower mobility carriers are accumulated by the gate at high voltages.

How universal is the observed critical density? When disorder strongly influences the transport, the critical density is found to change from sample to sample. However, in all the higher mobility samples that we measured, we observe a striking universality in the critical density. In figures 3b-e we show measurements on additional samples whose LAO layer thicknesses varied from 6 to 10 unit cells and whose mobilities varied by an order of magnitude, analyzed like in Fig. 3a. In all cases, we observed similar Hall and longitudinal resistance traces and comparable critical points, where the transport transitioned from one to two carriers (see supplementary A2). Whereas the critical gate voltage and the mobility at the transition, $\mu_C$, varies substantially between samples, the critical density in these high mobility samples is almost the same. Plotting the critical density vs. the critical mobility measured in several independent Hall bars on four different samples (Fig. 3f), we see that these densities have a universal value $n_C = 1.68 \pm 0.18 \times 10^{13} cm^{-2}$ independent of disorder, pointing to an intrinsic origin.

We argue that the transition occurs when the density exceeds a critical value at which new bands are populated. It is well established that interfacial electrons reside in the $t_{2g}$ conduction bands of STO[14,15,25,26]. In the bulk, these three bands are degenerate, having ellipsoidal Fermi surfaces centered at the $\Gamma$ point and oriented along the main lattice directions. At the interface this degeneracy is lifted due to tetragonal distortion and quantum confinement. Specifically, the $d_{XY}$ band, which has a larger effective mass in the z-direction than the $d_{XZ}$ and $d_{YZ}$ bands, splits toward lower energy[14] (Fig. 3g). Therefore below a critical carrier density all electrons are expected to reside in the $d_{XY}$ band and only one type of carriers exists in the system. At higher densities the Fermi energy enters the $d_{XZ}$ and $d_{YZ}$ bands, and as they get populated we expect to observe contributions from carriers of a new type.

The transition into the new bands is accompanied by a large change in the density of states (DOS), determined by the geometric mean of the effective masses in the two-dimensional plane, $\sqrt{m_x m_y}$. Since the $d_{XZ}$ and $d_{YZ}$ bands each has one light and one heavy mass in the plane whereas the $d_{XY}$ band has two light masses, the latter has the lowest DOS (Fig. 3h). Using ARPES measured effective masses[15], the combined DOS of the $d_{XZ}$ and $d_{YZ}$ bands is estimated to be $\approx 10$ times higher than that of the $d_{XY}$ band. Therefore, after raising the Fermi energy above the critical density we expect the newly occupied heavy bands to take up most of the added carriers while the density of the light $d_{XY}$ carriers to remain almost constant. This scenario perfectly fits our observations as plotted in Figs. 3a-e.

We can now address the observed universality of the critical density. Within the above band model the critical density is determined by the number of carriers that the $d_{XY}$ band can accommodate before the higher bands become available, $n_c = \Delta_E m_l / \pi \hbar^2$. Here $\Delta_E$ is the energy splitting between the $d_{XY}$ and the higher two bands, and $m_l$ is the light mass. Using the x-ray measured splitting[14], $\Delta_E = 50 meV$, and ARPES measured light mass[15], $m_l = 0.7 m_e$ ($m_e$ is the electronic mass), we estimate that $n_C = 1.46 \times 10^{13} cm^{-2}$, in good agreement with our experimentally determined value of $n_C = 1.68 \pm 0.18 \times 10^{13} cm^{-2}$. In general, $\Delta_E$ can have non-universal contributions from gate-induced electric fields, and a universal contribution from a charge or a dipole layer at the LAO/STO interface, as implied by recent measurements[20]. Our observation of a universal transition density suggests that an intrinsic mechanism, such as the proposed dipole layer, is the dominant component of the confinement potential at the interface.

The body of evidence presented above lends strong support to the interpretation that the observed transition occurs when d-bands of a different symmetry begin to be populated. It is interesting to see how this physics comes to bear on superconductivity in the system. Fig. 4 shows the longitudinal resistivity $\rho_{XX}$ measured down to $T = 50 mK$, divided by its normal-state value at $T = 0.38 K$, as a function of temperature and gate voltage. As reported earlier[3,7-9], we find that superconductivity is dome-shaped as a function of $V_G$. The superconducting transition temperature, $T_C$, first increases rapidly with $V_G$ up to a peak value, $T_C^{max} = 320 mK$, after which it slowly decreases with increasing $V_G$. Comparing these measurements to the critical density (dashed line in Fig.

4) we find that the latter corresponds to the peak of the dome. A similar correlation could also be deduced for published data[8] if we apply to it the analysis described above.

We now turn to discuss the relation between the above band picture and spin-orbit (SO) interactions. Earlier transport measurements[9,10] have pointed to a correlation between the dome-shaped variation of $T_C$ and the SO coupling scale in the system. At a certain density the apparent strength of SO rises sharply as a function of gate voltage, together with $T_C$, up to the peak of the superconducting dome[9]. Above optimal doping both the strength of SO and $T_C$ slowly decrease with increasing density[10]. So far, these observation have been interpreted to result from Rashba SO interactions that are induced by the electric field of the confining potential[9,10,24]. However, in a simple Rashba picture it is unclear why SO will suddenly increase at a certain density, nor why similar SO effects are observed in δ-doped STO[28], whose confining potential is symmetric.

We argue that the observed density dependence of SO can be naturally understood within the band picture discussed above. The key is to realize that the strongest SO effect in this system is the atomic spin-orbit (ASO) coupling[29,30]. ASO is especially prominent when the d-bands are degenerate, so that the coupling strongly mixes the orbitals and entangles the spin and orbital degrees of freedom. When we add ASO to the band calculation (Figs. 3i,j) we indeed see that its most dramatic effect is near the bottom of the $d_{XZ}$ and $d_{YZ}$ bands. Interestingly, we find that the calculated ASO strength as a function of energy (Fig. 3k, see caption for details) peaks around the critical point in a striking resemblance to the experimental observation, suggesting that intrinsic ASO at the band transition is responsible for the strong SO observed in this system. In the supplementary material (section A1) we further show that the intimate connection between the critical transition and ASO may also explain other observations in this system, most notably the low carrier densities measured by Shubnikov-de-Haas[11,12,31-33].

The ASO, together with the asymmetric confinement potential on the surface can also give rise to small Rashba-like SO terms (supplementary A4). In presence of such coupling, the anisotropy of transport with in-plane magnetic field can serve as a probe of the orbital symmetry of the conduction bands[34]. Other measurements that we will report in details elsewhere show that such anisotropy switches in perfect synchrony with the transition from one to multiple bands. Below the critical density we find a small anisotropy oriented along the magnetic field, consistent with a circular $d_{XY}$ band, whereas above the critical point the anisotropy is large and is oriented along the crystal axes, as expected from elliptical $d_{XZ}$ and $d_{YZ}$ bands.

In summary, we observe that the two-dimensional system formed at the LAO/STO interface undergoes a transition which reshapes its core transport properties above a critical carrier density. We show that this transition appears at a universal critical density over a large variety of samples, and that it can be naturally understood as a change from single to multiple d-band population. We find that prominent ASO interactions near this transition can account for the peak of SO strength observed experimentally. We further observe a similar correlation between this transition and the peak of superconducting $T_C$, which may suggest that the population of the $d_{XZ}$ and $d_{YZ}$ bands at this critical density plays an important role in the emergence of superconductivity as well as possibly other correlated states in this system. These observations demonstrate that degeneracies of bands with different symmetries lie at the root of the fascinating transport properties observed so far at the interface of the insulating oxides.

**Acknowledgements:** We would like to acknowledge A. D. Caviglia, S. Gariglio, A. Fete and J. –M. Triscone for the samples, fruitful discussions, and useful comments on the manuscript. We benefited greatly also from discussions with E. Berg, Y. Gefen, D. Goldhaber-Gordon, Y. Meir, Y. Oreg, D. Shahar, A. Stern and A. Yacoby. S.I. acknowledges the financial support by the ISF Legacy Heritage foundation, the Minerva foundation, the EU Marie Curie People grant (IRG) and the Alon fellowship. S.I. is incumbent of the William Z. and Eda Bess Novick career development chair. E.A. acknowledges the financial support by the ISF foundation. E.A. is incumbent of the Louis and Ida Rich Career Development Chair.

**Figure 1: Observation of a critical transition in transport at the LAO/STO interface**. a) Measured Hall resistance, $\rho_{XY}$, vs. magnetic field, $B$, for various back-gate voltages, $V_G$, in 20V steps. At a critical value, $V_C = 40V$, a transition is observed between two different types of $B$ dependences. b) Zoom-in on the low-field region of panel a. The low-field slopes reverse their gate dependence at $V_C$ (inset): Below $V_C$ they decrease with increasing gate-voltage, they saturate near $V_C$, and then slightly increase. c) Corresponding Hall coefficient, $R_H = \frac{\rho_{XY}}{B}$, plotted vs. $B$ for the same $V_G$'s as in panel a. Below $V_C$, $R_H$ is roughly independent of $B$, but above $V_C$ it peaks at an intermediate field and falls off at high fields. Indicated for the highest $V_G$ is the position of the peak, $B_P$, and the falloff field, $B_W$, which is determined from the deflection point of the curve. These characteristic fields depend on $V_G$. d) Measured longitudinal resistivity, $\rho_{XX}$, normalized by its value at $B = 0$, $\rho_{XX}^0$, plotted as a function of $B$ for the same $V_G$'s as in panel a.

**Figure 2: Scaling of transport curves above and below the critical gate voltage, $V_C$.** The curves in Figs. 1c-d separate into two groups which scale differently. a) Below $V_C$, the $R_H$ traces merge by a mere shift along the y-axis b) The normalized $\rho_{XX}$ curves below $V_C$ are already self-similar. c) Above $V_C$, the $R_H$ curves differ substantially in their characteristic fields, but when plotted vs. $B$ divided by the characteristic width, $B/B_W$, after subtracting a y-offset and scaling of their height, all of them collapse perfectly. Lorentzian fit (dashed line). d) The $\rho_{XX}$ curves above $V_C$, plotted as a function of $B/B_W$, where $B_W$ is taken from the $R_H$ curve at the same gate voltage. This scaling yields perfect collapse also of the $\rho_{XX}$ traces, showing the strong correlation between $\rho_{XX}$ and $R_H$. Lorentzian fit (dashed line) e) $B_W$ plotted vs. carrier density, $n_{total}$, obtained from the high-field Hall data (the grey dashed line is a guide to the eye). $B_W$ appears to diverge at a critical density, $n_C$. f) This critical density is estimated from the x-intercept of the plot of the inverse field, $1/B_W$, vs. density (the dashed line is a linear fit to the low-density data).

**Figure 3: Universality of the critical density and its energy bands origin.** a) Densities extracted from the measured Hall coefficient ($n = 1/eR_H$, $e$ is the electronic charge) at B=0T (red) and at B=14T (blue) as a function of $V_G$. Within a simple model

(see text) the former reflects the density of high-mobility carriers, $n_{hi}$, and the latter the total density, $n_{total}$. Below $V_C$ (dashed vertical line), $n_{total} \approx n_{hi}$ and both increase with gate voltage up to the critical value, $n_C$, (dashed horizontal line) reached at $V_C$. Above $V_C$, $n_{hi}$ saturates whereas $n_{total}$ continues to increase. b) to e) Similar analysis on additional samples differing by their LAO layer thickness and growth temperatures. f) Collection of the critical densities measured with several independent Hall bars on four samples (different colors), plotted as a function of the electron mobility at the critical point, $\mu_C$. These critical densities show a universal value, $n_C = 1.68 \pm 0.18 \times 10^{13}\ cm^{-2}$, independent of the mobility and LAO thickness. g) The d-orbital energy bands of STO near its interface with LAO. Quantum confinement in the z-direction moves the $d_{XY}$ band, which is heavy in this direction, to energy lower by $\Delta_E$ than the $d_{XZ}$ and $d_{YZ}$ bands, which are light in this direction. h) The corresponding schematic density of states (DOS) as a function of the energy. The $d_{XY}$ band has ~10 times smaller DOS (red) than the combined DOS of the $d_{XZ}$ and $d_{YZ}$ bands (blue) (see text). i) Calculated energy bands including atomic spin-orbit interactions (ASO), which open gaps at all degeneracy points of the d-orbitals. Specifically, at the $\Gamma$ point the bands split by $\Delta_{ASO}$, the parameter in the atomic spin orbit Hamiltonian, and mix into two separate branches with light and heavy character (supplementary A1). j) Calculated DOS with ASO, showing that the sharp jump of panel h is smeared over a $\Delta_{ASO}$ scale. k) The strength of ASO, $\langle L \cdot S \rangle$, integrated over the Fermi surface of the first two bands and plotted as a function of the energy (green). Notably, ASO is strongly peaked near the critical point, where its effect is most prominent due to orbital degeneracy. The averaged $\langle L \cdot S \rangle$ in the third band is shown in purple. The figure shows the absolute value of this quantity, which has the opposite sign than that in the first two bands. Clearly, ASO is also sharply peaked near the bottom of the third band. For typical carrier densities in this system, however, the Fermi energy is always below the bottom of this band.

**Figure 4: Correlation between superconductivity and the critical density.** $\rho_{XX}$ measured vs. temperature and gate voltage, divided by its normal-state value at T=0.38K. Red and blue colors correspond to the normal and superconducting states. The superconducting transition temperature, $T_C$, defined by $\rho_{XX}(T)/\rho_{XX}(T = 0.38K) = 1/2$,

has a dome shape as function of $V_G$. The critical density (white dashed line) is found to be at the peak of the dome.

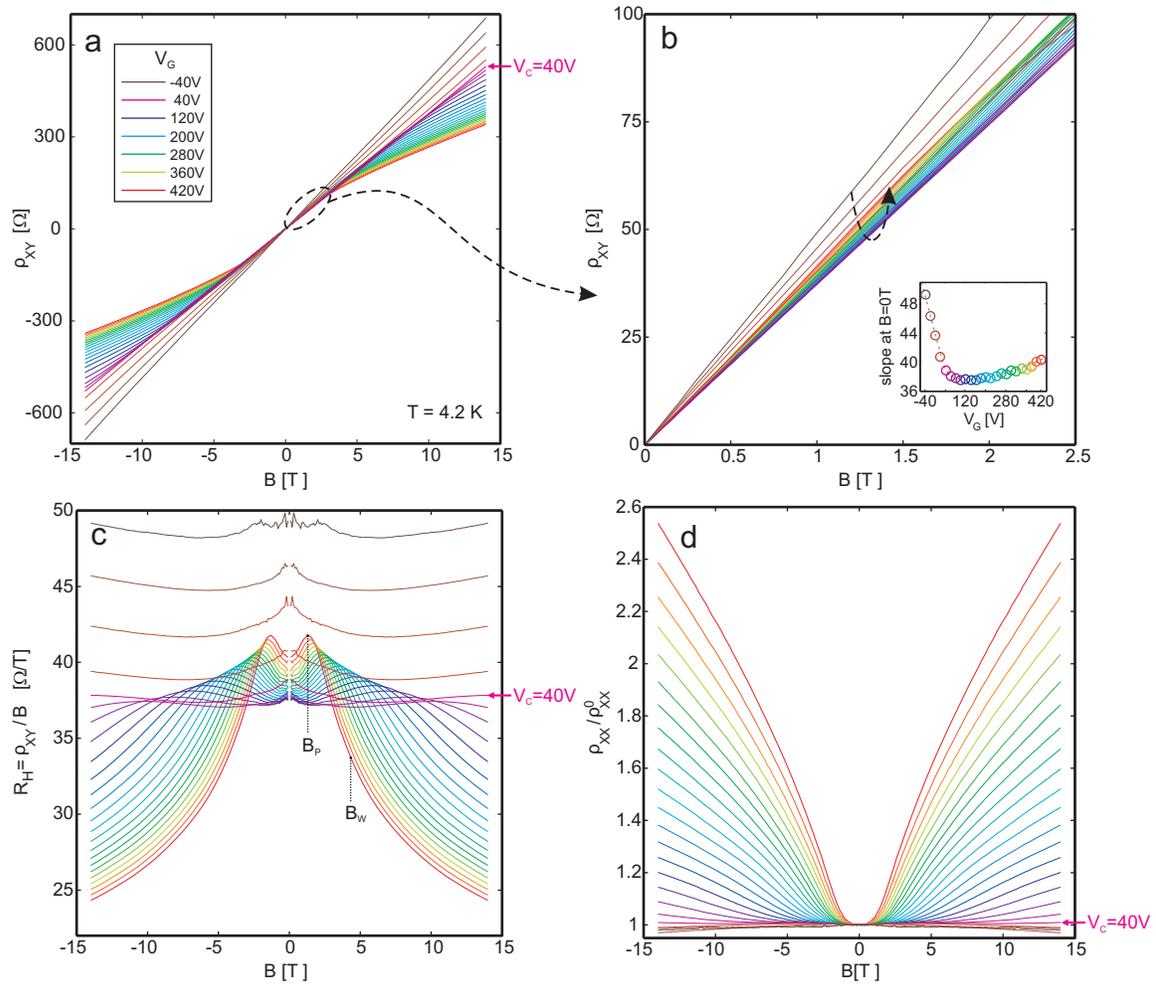

**Figure 1**

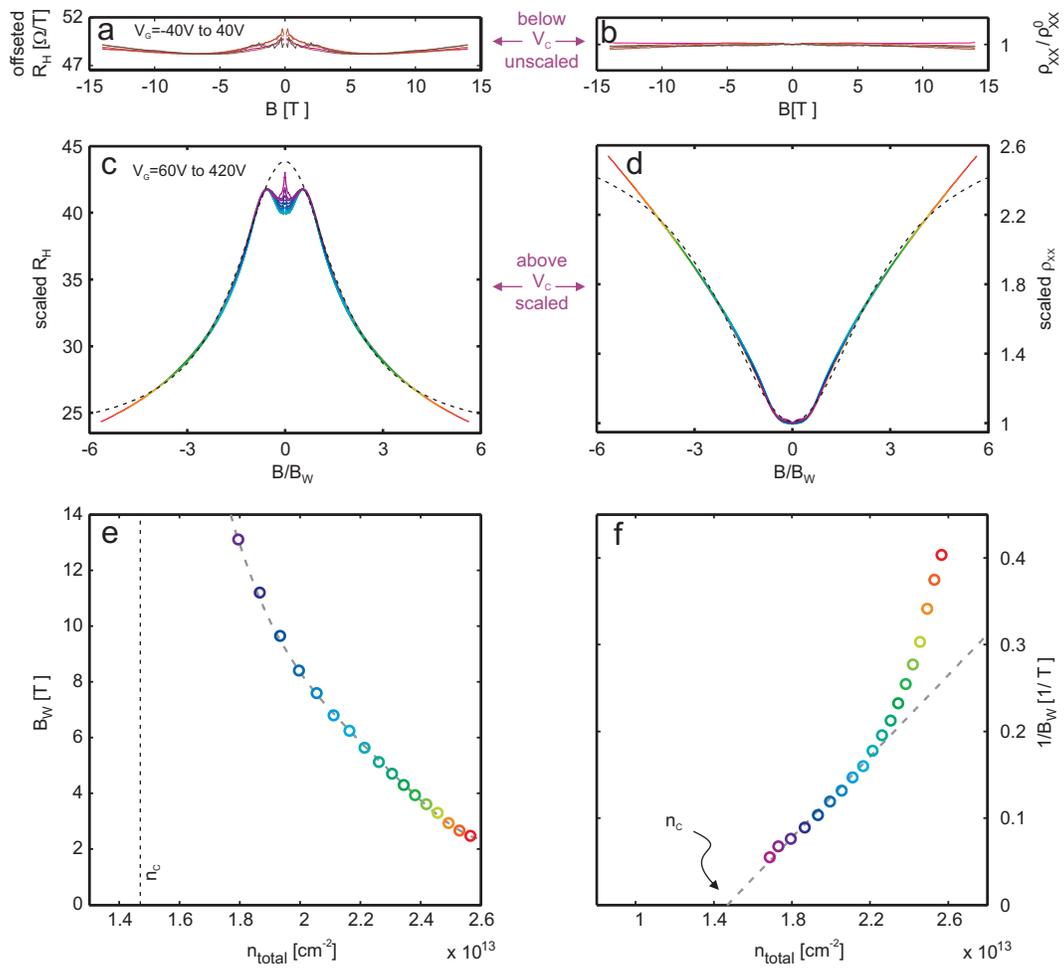

**Figure 2**

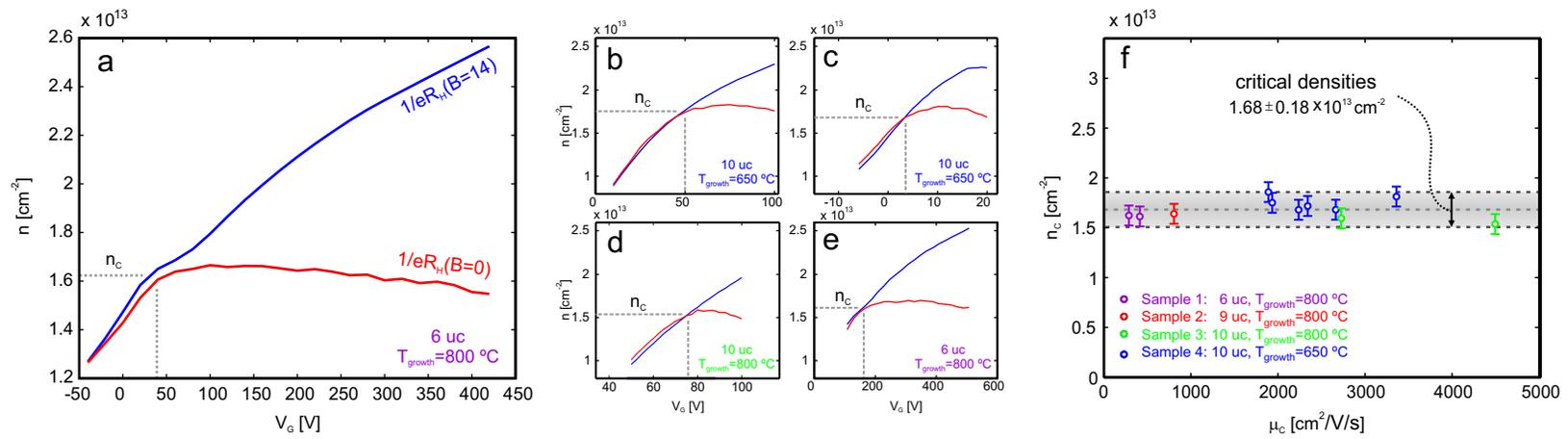
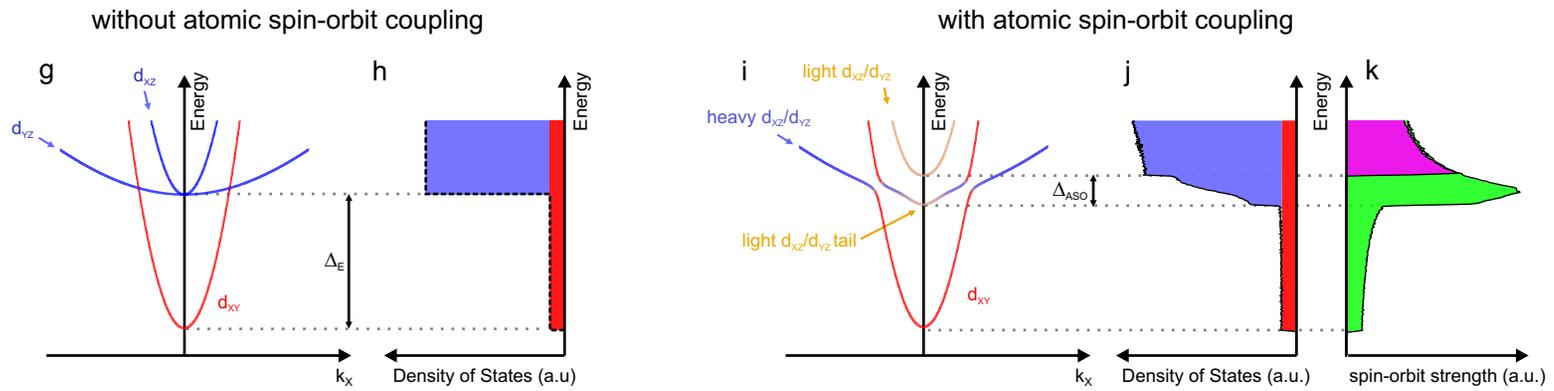

**Figure 3**

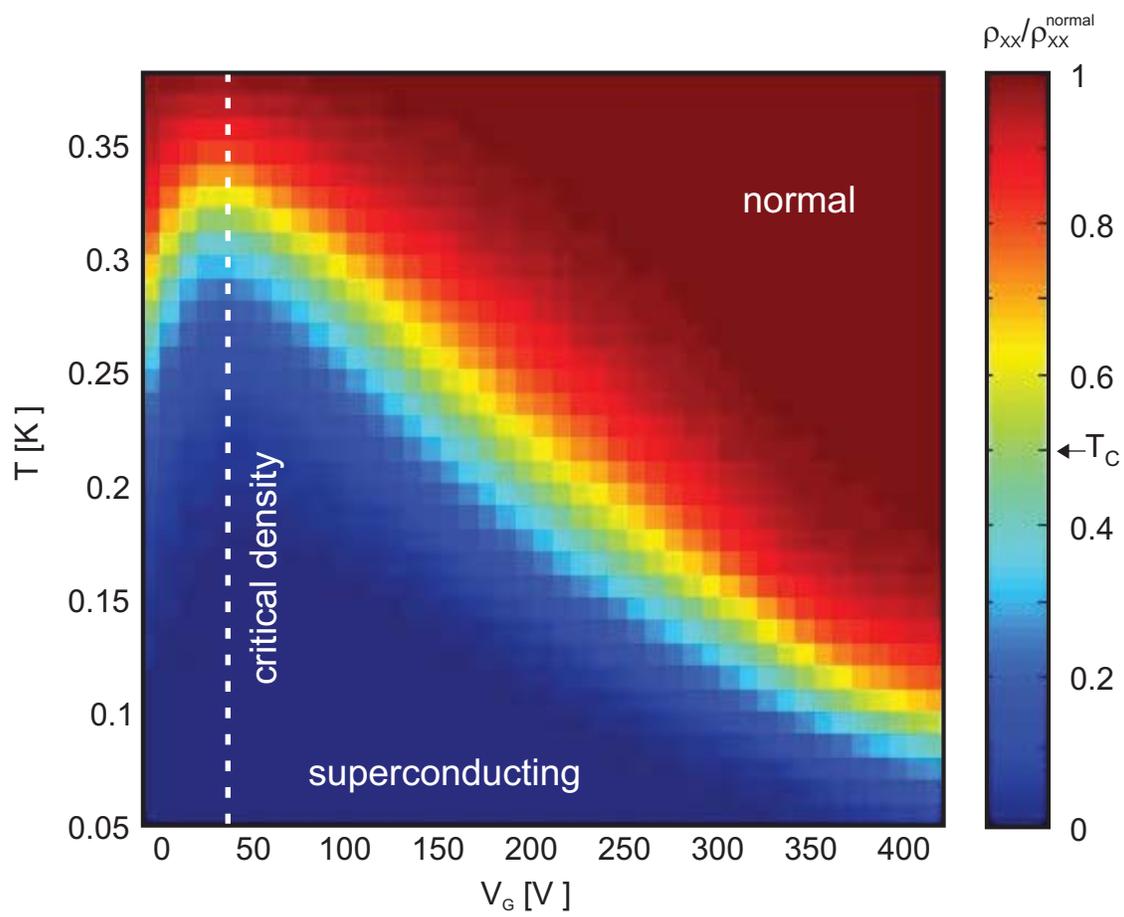

**Figure 4**

# SUPPLEMENTARY INFORMATION

## A Universal Critical Density Underlying the Physics of Electrons at the LaAlO$_3$/SrTiO$_3$ Interface

Arjun Joshua, S. Pecker, J. Ruhman, E. Altman and S. Ilani

*Department of Condensed Matter Physics, Weizmann Institute of Science, Rehovot 76100, Israel.*

A1. **The band structure with atomic spin-orbit interactions and its relation to the observed transport properties.**

A2. **Data from additional samples.**

A3. **The two-band approximation to the transport above the critical density.**

A4. **Details of the calculation of the energy bands with atomic spin-orbit interactions**

A5. **Sample fabrication**

## A1. The band structure with atomic spin-orbit interactions and its relation to the observed transport properties.

In the first part of the main text we used the simplest band model that captured the essential features of the critical transition. We then showed that the inclusion of atomic spin-orbit (ASO) interactions can further explain the strong correlation between the critical point (the bottom of the $d_{XZ}$ and $d_{YZ}$ bands) and strong spin-orbit interactions, as observed experimentally. In this section we demonstrate that the refined band structure with ASO has additional features that may explain other surprising observations in the LAO/STO system. For example, we argue that such a model can naturally explain why carrier densities extracted from Shubnikov-de-Haas oscillations are so much lower than those determined by Hall measurements[1-5].

Figure S1 shows the energy bands calculated with and without ASO (colored and grey lines correspondingly). The details of the calculations are explained in section A4 below. As can be seen in the figure, the effect of ASO is strongest near band degeneracy points.

At these points ASO hybridizes the electronic states and opens energy gaps in the spectrum. Such degeneracies exist between the $d_{XZ}$ and $d_{YZ}$ bands at the $\Gamma$ point of the Brillouin zone, and between the $d_{XY}$ and the other two bands at their crossing points. Since the heavy band is very shallow, these three types of crossings appear close in energy, around the transition point.

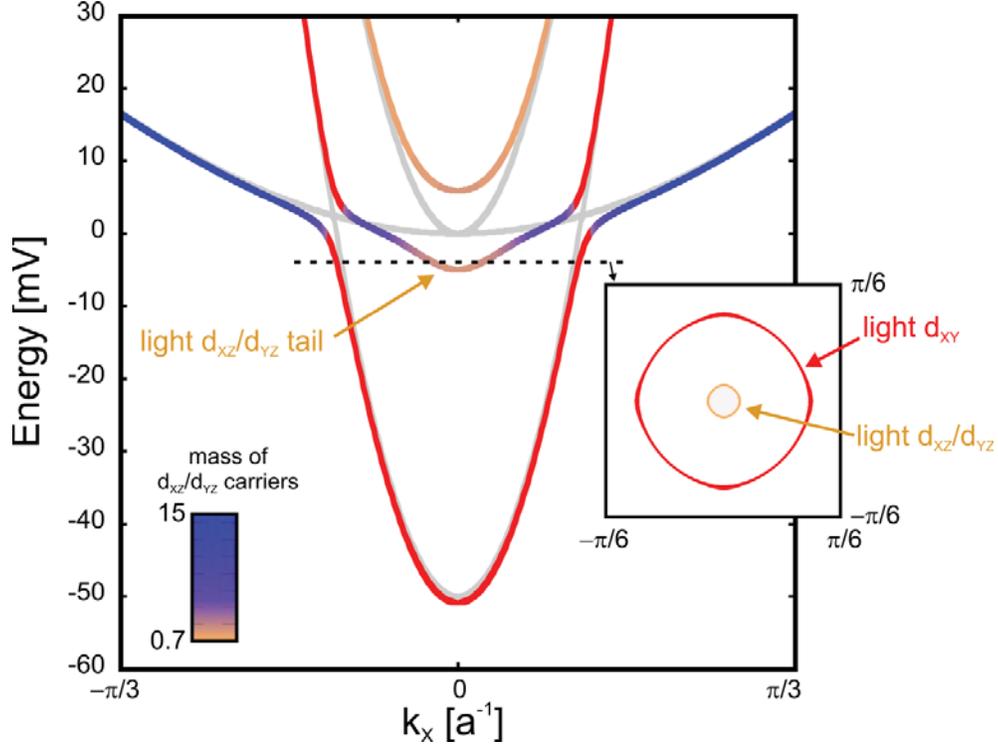

**Figure S1: The band structure at the LAO/STO interface with atomic spin-orbit (ASO) interactions.** The main panel shows the energy bands without ASO (gray) and with ASO (colored) calculated with the ARPES measured[6] values for the light and heavy masses, $m_l = 0.7 m_e$ and $m_h = 15 m_e$ ($m_e$ is the electronic mass), an energy splitting of $\Delta_E = 50 meV$, and taking the strength of ASO to be $\Delta_{ASO} = 10 meV$. In the calculation with ASO the states are colored according to their orbital content and effective mass: States whose $d_{XY}$ content is larger than 20% are colored red. The remaining $d_{XZ}/d_{YZ}$ states are colored according to their effective mass, ranging from brown (light mass) to blue (heavy mass) (see colorbar). The mass plotted is that along the x direction, $m^* = \hbar^2 k_x/(dE/dk_x)$, but qualitatively similar behavior is obtained when we plot the mass averaged over the entire Fermi surface. ASO modifies the bands mostly around their degeneracy points. One consequence of that is that the heavy (blue) $d_{XZ}/d_{YZ}$ carriers become light (brown) at the bottom of these bands. Inset: The Fermi surfaces 1meV above the bottom of the $d_{XZ}/d_{YZ}$ bands, showing large circular Fermi surface of light $d_{XY}$ carriers and a small circular Fermi surface of light $d_{XZ}/d_{YZ}$ carriers.

ASO has two important consequences that significantly affect the transport in this system:

**1. Spin is strongly coupled to the orbital momentum near the critical density.** Around this point the $d_{XY}$, $d_{XZ}$ and $d_{YZ}$ bands cross, allowing the spin-orbit coupling to strongly hybridize them and form superpositions with well-defined atomic orbital momenta. For example, the degenerate $d_{XZ}/d_{YZ}$ states at the $\Gamma$ point hybridize to form the $d_{\mp} = d_{XZ} \pm id_{YZ}$ states, with a well-defined projection of the orbital momentum along the z direction, $m_z = \pm 1$. At the same time, the spin becomes strongly coupled to this orbital momentum. Since the orbital momentum has a preferred axis of orientation (e.g. the z-axis for the example above), the spin is also preferably aligned along this axis, and an energy of the order of $\Delta_{ASO}$ is required to polarize it away from this axis.

When the Fermi energy is increased above the degeneracy points, the band splittings due to the different dispersion of the $d_{XY}$, $d_{XZ}$ and $d_{YZ}$ orbitals become increasingly dominant over the ASO coupling and the effect of the latter decays. In Fig. 3k of the paper the enhancement and subsequent decay of the characteristic spin-orbit scale is seen from the variation with the Fermi energy of the expectation value $\langle L \cdot S \rangle$, integrated around the Fermi surface.

Further refinement to this picture is given by the transverse confining potential, which breaks the inversion symmetry perpendicular to the plane. This allows for a small tunneling matrix element between d-orbitals of different symmetries on neighboring atoms. Combined with the strong ASO coupling, this additional tunneling leads to a small, Rashba-like splitting of the energy bands away from the $\Gamma$ point. This effective Rashba coupling is a direct consequence of the ASO coupling and is thus also peaked near the critical density. However, since it relies on small, field-induced hopping elements, its effect would be much weaker than that of the ASO coupling.

**2. The heavy $d_{XZ}/d_{YZ}$ carriers become light at the bottom of their bands.** The $d_{XZ}/d_{YZ}$ carriers play a central role in the observations reported in this paper. These carriers have a much larger in-plane mass than the light $d_{XY}$ carriers. Therefore their population is associated with a large jump in the DOS, which is consistent with the transition that we measure in transport. In a simple band model without ASO this jump is abrupt, occurring when the Fermi energy crosses the bottom of the $d_{XZ}/d_{YZ}$ bands. ASO, however, modifies this picture by drastically changing the mass of the $d_{XZ}/d_{YZ}$ carriers near the bottom of their bands.

To demonstrate this effect we have colored the eigenstates in Fig. S1 according to their orbital content and effective masses. The $d_{XY}$ carriers are colored red, and the $d_{XZ}/d_{YZ}$ are colored by a continuum of colors that represent their effective mass, ranging from brown (light mass) to blue (heavy mass) (see caption for details). At high energies the $d_{XZ}/d_{YZ}$ states have indeed a heavy mass (blue). However, close to the bottom of the band we can clearly identify a low-energy tail in which the $d_{XZ}/d_{YZ}$ carriers are light. This effect is a robust consequence of the hybridization gap created by ASO at the degeneracy point between the $d_{XZ}/d_{YZ}$ orbitals. While the amount of light carriers contained in this low-energy tail is proportional to the strength of ASO, the light mass obtained at its bottom is always the same. In fact, it is straight forward to show that this mass is approximately twice the light mass, $m^* = 2(m_h^{-1} + m_l^{-1})^{-1} \approx 2m_l$ (where $m_l$ and $m_h$ are the light and heavy masses in the absence of ASO). In contrast to the case without ASO, here the mass of the $d_{XZ}/d_{YZ}$ carriers depend on the Fermi energy, evolving continuously from a light mass at the bottom of the band to a heavy mass at higher Fermi energies.

To further demonstrate the light nature of the $d_{XZ}/d_{YZ}$ carriers near the bottom of the band, we plot in the inset to Fig. S1 the Fermi surfaces at low band filling. In addition to the circular $d_{XY}$ Fermi surface we see a small, circular, Fermi pocket that corresponds to the light $d_{XZ}/d_{YZ}$ carriers. This is in sharp contrast to the picture without ASO where the $d_{XZ}/d_{YZ}$ Fermi surfaces are elliptical and heavy even at the bottom of their band. While the actual size of ASO in this system is yet to be determined experimentally, from transport experiments[7,8] and ab-intio calculations[9] we expect it to be in the range $\Delta_{ASO} \approx 10 - 25$meV. For such ASO strengths the density of light carriers that the $d_{XZ}/d_{YZ}$ band can accommodate before it becomes heavy is about $1.5 - 5 \cdot 10^{12}$cm$^{-2}$.

The existence of light $d_{XZ}/d_{YZ}$ carriers has important implications for the transport in this system. On Hall measurements we expect the effect of these carriers to be rather small. The Hall resistance measures the parallel addition of these carriers and the $d_{XY}$ carriers. Since the effective masses of these two types of carriers are comparable and thus also their mobilities, even when these two types of carriers coexist, their combined Hall resistance would behave as the Hall resistance of a single carrier-type with the total density of both. A dramatically different result will occur when one measures the Shubnikov-de-Haas oscillations. These oscillations measure the individual bands independently, giving a different oscillation frequency for each carrier type, which is directly proportional to their individual densities. Since the density of the light $d_{XZ}/d_{YZ}$ carriers is significantly smaller than that of the $d_{XY}$ carriers, it would result in substantially different oscillation period as a function of 1/B. In fact, all Shubnikov-de-Haas measurements in LAO/STO interfaces[1,2] as well as those in δ-doped STO[3-5] have robustly observed a pocket of light mass carriers with densities in the range $1.3 - 4.8 \cdot 10^{12} \text{cm}^{-2}$. These measurements extracted a mass of $m^* \approx 1.2 - 1.4 m_e$, which is twice that measured by ARPES[6,10]. The factor two discrepancy between these two measurement techniques is in fact consistent with the ASO model; ARPES experiments, done on a bare STO surface with much higher electron density, determines the masses from high energies and thus yields the bare light mass, whereas transport is sensitive to the low energies dispersion which is renormalized by ASO. So far the low density pockets measured by Shubnikov-de-Haas were explained by higher subbands of $d_{XY}$ carriers. However, as we see here, the effect of ASO provides an alternative, robust, and intrinsic mechanism for the existence of such light carriers, of the $d_{XZ}/d_{YZ}$ type.

There are two important features of the light $d_{XZ}/d_{YZ}$ carriers that set them apart from light $d_{XY}$ carriers at higher subbands. First, in the ASO model the appearance of the light pocket is intimately tied to the bottom of the $d_{XZ}/d_{YZ}$ bands, namely to the critical density. This is in contrast to higher $d_{XY}$ subbands whose energies do not have to coincide with the $d_{XZ}/d_{YZ}$ band bottom. Secondly, a scenario that includes several $d_{XY}$ subbands would involve several discrete transitions as a function of the Fermi energy, e.g. a transition from a single light carrier type at the lowest $d_{XY}$ subband to two carrier types in the two lowest $d_{XY}$ subbands, and finally to three carriers types that include also the heavy $d_{XZ}/d_{YZ}$ carriers. In contrast, in the ASO model when the $d_{XZ}/d_{YZ}$ start populating there are always only two carrier types. At low filling these are the $d_{XY}$ and light $d_{XZ}/d_{YZ}$ carriers and at high filling they are the $d_{XY}$ and heavy $d_{XZ}/d_{YZ}$ carriers.

Although the transition between these two cases is rather sharp as a function of increasing Fermi energy, it is a *continuous* crossover. The strong two-band signatures that we observed in the paper should appear only when heavy carriers are populated. However, if this model is correct, we might expect to see weak signatures of the light mass tail even below the critical density and these signatures should evolve *continuously* to the strong two-band features above the critical density. In the next section, we present detailed Hall resistance data that complements the one shown in the main text and analyze it in light of these predictions.

**A2.     Data from additional samples.**

Figure 3f of the main text collected data from four LAO/STO samples that differed in the number of LAO unit cells and in their mobilities. All these samples showed a critical transition at the same universal density. In Fig. 1 we showed the detailed Hall traces of sample 1 with 6 unit cells of LAO. In this section we present the detailed Hall traces of samples 2 - 4 and analyze them in light of the ASO model described in the previous section.

In the main text we demonstrated that both the Hall resistance, $\rho_{XY}$, and the Hall coefficient, $R_H = \rho_{XY}/B$, show clear signatures of the critical density. To make the identification of this density even more intuitive we plot in Fig. S2a a related quantity, the effective density, $\tilde{n} = 1/(e \cdot d\rho_{XY}/dB)$, determined from the derivative of $\rho_{XY}$ with respect to $B$. Compared to $R_H$, this derivative is a local quantity and thus accentuates the magnetic field-dependent features in the curves. For each sample, we plot in Fig. S2a, $\tilde{n}$ as a function of $B$ for various gate voltages, $V_G$. Notably, in all samples we can identify a clear critical transition density when we go from low $V_G$ (brown traces) to high $V_G$ (red traces), as is explained below.

There are two sharp criteria to identify the critical density, $n_c$. The first is that below $n_c$ the Hall slopes (and correspondingly $\tilde{n}$) are identical at zero and high field, whereas above $n_c$ they are different. The second is that once $n_c$ is crossed, the low field density, $\tilde{n}(B = 0T)$, remains stuck at $n_c$. Clearly, all the samples in Fig. S2a exhibit these two changes around the universal value of $n_c$ (dashed line in the figure). Below $n_c$, in all the curves $\tilde{n}(B = 9T)$ is roughly equal to $\tilde{n}(B = 0T)$ whereas above $n_c$ the former becomes larger than the latter, and their difference grows with increasing $V_G$. Furthermore, once $\tilde{n}(B = 0T)$ reaches $n_c$ it remains stuck at this value although the total density continues to rise, as is evident from the crowding of curves near $n_c$ at $B = 0T$. These two changes are seen more directly when we plot $\tilde{n}(B = 0T)$ and $\tilde{n}(B = 9T)$ as a function of $V_G$ for all samples (red and blue traces respectively, in Figs. 3b-3e).

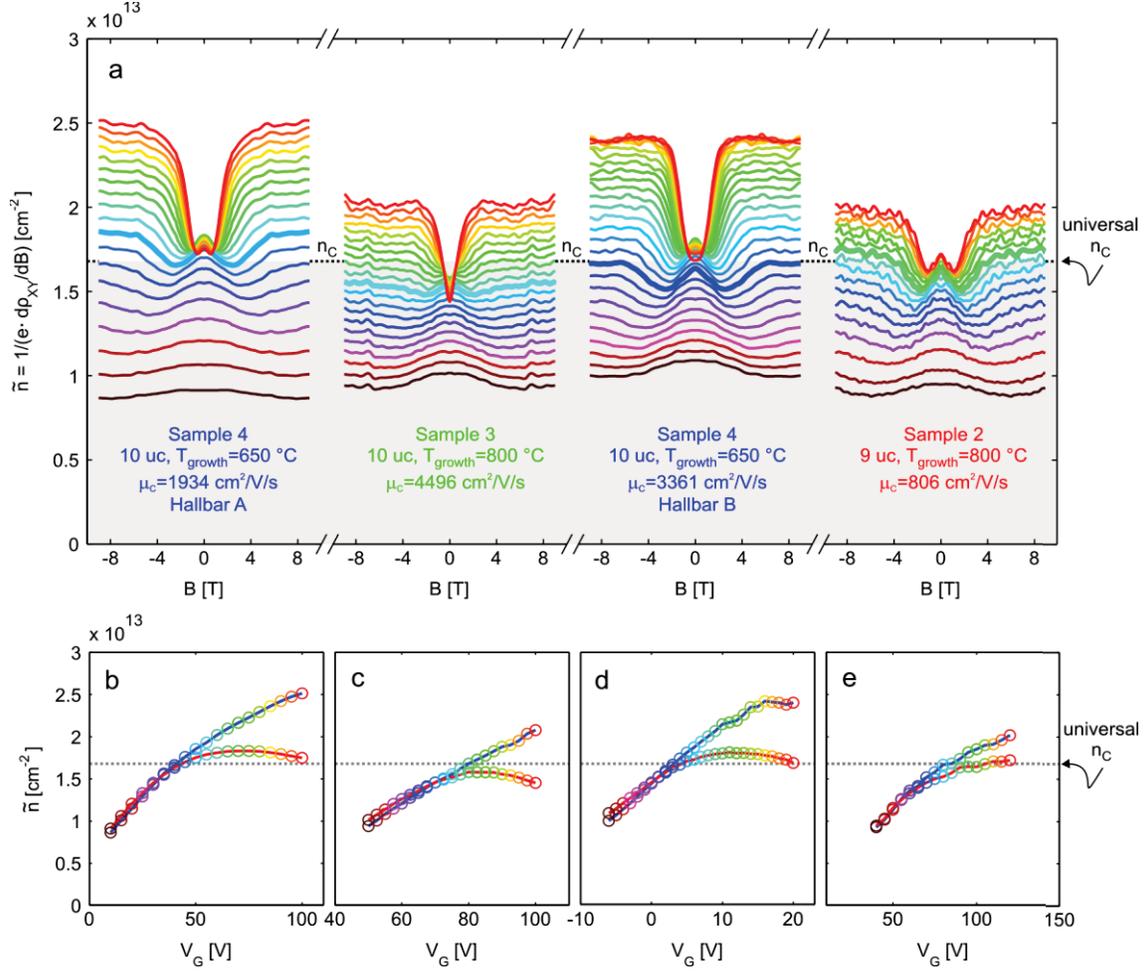

**Figure S2: The universality of the critical density and of Hall traces across multiple samples.** The figure shows the data for samples 2 – 4 (Similar data from sample 1 is shown in the main text). a) The effective density, $\tilde{n} = 1/(e \cdot d\rho_{XY}/dB)$, calculated from the derivative of the measured Hall resistance with respect to magnetic field, $d\rho_{XY}/dB$, is plotted as a function of $B$, for various gate voltages, $V_G$ (brown to red go from the lowest to the highest $V_G$). For each sample we indicate the LAO layer thickness, growth temperature and measured mobility at the critical transition (see main text). The dashed black line marks the universal critical density, $n_c = 1.68 \cdot 10^{13}\ cm^{-2}$, derived in the main paper. In all samples, independent of mobility and LAO thickness, we observe a clear change of the characteristic traces around the universal density (see text). b) - e) Corresponding plots of the effective density at zero field, $\tilde{n}(B = 0T)$ [red line], and high field, $\tilde{n}(B = 9T)$ [blue line], as a function of $V_G$ for the samples in panel a. For all samples these two density traces depart near the universal critical density, marked by a grey line in the figures. Above this density, $\tilde{n}(B = 9T)$ continues to rise whereas $\tilde{n}(B = 0T)$ remains roughly constant.

Careful examination of the curves reveals that at intermediate magnetic fields there are fine features, which are not captured by the above analysis. These are apparent as small dips in $\tilde{n}$, which are equivalent to the peaks of $R_H$ at $B = B_P$ shown in Fig. 1c for sample 1. The physics of these dips in $\tilde{n}$ (peaks in $R_H$) will be described in a future publication. Here, however, we want to comment on their relation to the multiple-band physics. Above $n_c$ these dips show a strong correlation with the signatures of multiple bands. As a function of magnetic field these dips always appear right before the sharp rise in $\tilde{n}$, associated with the multiple bands. The magnetic field position of this rise varies substantially with gate voltage, however, as can be seen from Fig. S2a the dips in $\tilde{n}$ always perfectly track the position of this rise. The exact same behavior was observed for sample 1 in the main text, where we showed that peaks in $R_H$ and the $R_H$ falloff associated with multiband physics scale together in magnetic field (Fig. 2c).

Interestingly, we can see in Fig. S2a that the dips in $\tilde{n}$ appear also below $n_c$. Moreover, if we track the magnetic field position of these dips we see that they evolve continuously across the transition. These observations hint that even below $n_c$ there is a mild multiple-band behavior which evolves continuously into the strong multiple-band behavior commencing at $n_c$. One might expect weak multiple-band signatures from higher $d_{XY}$ subbands, however, since these subbands are independent of the $d_{XZ}/d_{YZ}$ bands that are responsible for the strong multiple-band behavior above $n_c$, generally the former would not continuously evolve to the latter. A much more natural explanation for the above observation is given by the ASO model described in the previous section. Within this model the mass of the $d_{XZ}/d_{YZ}$ carriers changes rapidly but continuously from light to heavy as a function of the filling of these bands. It is thus straightforward to assign the weak multi-band signatures below $n_c$ to the light $d_{XZ}/d_{YZ}$ carriers and the strong multi-band signatures above $n_c$ to the heavy $d_{XZ}/d_{YZ}$ carriers at higher filling. Within this model the evolution would indeed be sharp yet continuous as we see in our measurements.

### A3. The two-band approximation to the transport above the critical density.

The simplest model for understanding the behavior of the transport above the critical density is the two-band approximation[11]. This model assumes that two types of carriers are conducting in parallel, each having its own density and mobility. It also assumes that the parameters of these bands are independent of $B$. This model is often used in the LAO/STO literature to extract the mobilities and densities of two carriers when Hall resistance traces are S-shaped. In this section we show that this model should be used with caution. While it can capture some basic features of the data, it can also lead to large mistakes in some of the extracted parameters. In the main text we were therefore careful to use only the robust predictions of this model that are insensitive to its assumptions which might be violated.

Assuming two parallel conducting bands whose densities and mobilities are $n_1, \mu_1, n_2, \mu_2$ the model predicts that the Hall coefficient of the combined system, $R_H$, and its longitudinal resistance, $\rho_{XX}$, should have Lorentzian shapes as a function of $B$:

$$R_H = R_\infty + \frac{R_0 - R_\infty}{1+(B/B_W)^2}, \quad \rho_{xx} = \rho_\infty + \frac{\rho_0 - \rho_\infty}{1+(B/B_W)^2} \qquad \text{(Eq. 1)}$$

where the asymptotic values of these two quantities at zero and infinite fields can be written in terms of the parameters of the bands as:

$$R_0 = \frac{n_1 \mu_1^2 + n_2 \mu_2^2}{(n_1 \mu_1 + n_2 \mu_2)^2}, \quad R_\infty = \frac{1}{(n_1 + n_2)} \qquad \text{(Eq. 2)}$$

and:

$$\rho_0 = \frac{1}{n_1 \mu_1 + n_2 \mu_2}, \quad \rho_\infty = \frac{(n_1 \mu_2 + n_2 \mu_1)}{(n_1 + n_2)^2 \mu_1 \mu_2} \qquad \text{(Eq. 3)}$$

The model further predicts that the Lorentzians of both $R_H$ and $\rho_{xx}$ should have the same characteristic width in magnetic field:

$$B_W = \frac{n_1\mu_1 + n_2\mu_2}{(n_1 + n_2)\mu_1\mu_2} \qquad (\text{Eq. 4})$$

In Figs. 2c and 2d in the main text we compare the measured $R_H$ and $\rho_{xx}$ traces at gate voltages above the critical density with a Lorentzian fit and indeed find a reasonable agreement with this functional form. Moreover, as is predicted by the model, we see that $R_H$ and $\rho_{xx}$ share the same characteristic field scale, $B_W$. We demonstrated this by showing that if for every gate voltage we use the characteristic field with which we scaled $R_H$ to scale the $\rho_{xx}$ at the same gate voltage, we get that all the $\rho_{xx}$ traces perfectly collapse onto a universal curve. Thus, overall it seems like this simple model can capture the basic features in our data.

However, if we carefully examine the fits of the model we can clearly identify systematic deviations. For example, we see that at low field both $R_H$ and $\rho_{xx}$ deviate from Lorentzians – the former peaks at a finite field $(B_P)$ and not at zero, and the latter is extremely flat up to $B_P$, both deviating from the simple Lorentzian dependence. Clear deviations are also observed at high fields, especially in $\rho_{xx}$, which shows a linear increase as a function of $B$ up to the highest field in our measurements, inconsistent with the Lorentzian shape. One might consider improving the model by implementing more realistic bands: three bands instead of two, elliptical Fermi surfaces instead of spherical, and even including the effect of ASO. We found, however, that even when we include these details, the above deviations could still not be explained.

In the literature the above details are often ignored, and a rough fit to kinked Hall curves is used to extract two densities and two mobilities for the carriers. Such fits can give apparently reasonable values of mobilities and densities, especially at high total densities. We find, though, that using this model brute-force to systematically study the gate dependence of the extracted parameters leads to unphysical results. For example, we get that the total density in the bands decreases with increasing gate voltage, which is clearly wrong. The most significant error in the model is in its prediction of the crossover field. As can be seen from Eq. 4, when the density of the second band, $n_2$, is low (as is the case near the critical point) the model predicts a crossover field at $B_W \approx 1/\mu_2$, the inverse of the lower mobility of the two bands, whereas in our data we consistently find that the crossover appear at $B_W \approx 1/\mu_1$, the inverse of the higher mobility of the two bands.

In a future publication we will describe a more elaborate model which we believe can account for the above observations. Here, however we do want to note that the main problem of the simple two-band model is its assumption that the parameters of the bands are independent of magnetic field. As a result, one has to be cautious when using this model, and extract from it only robust parameters that are insensitive to this assumption. The parameters that are most sensitive to this assumption are the mobilities of the two bands, which are directly affected by the inconsistencies between the model and the data in the value of $B_W$, described above. Thus, imposing the two band model on the data would often result in large errors in the extracted mobilities. In the main text we were therefore careful not to extract the mobilities, but extracted only the densities which are less sensitive to the errors of the model. Specifically, to extract the densities we have used in the main text two identities: The first relates the Hall coefficient at large fields with the total density, $R_\infty = 1/(en_{total})$. As can be seen from Eq. 2 above, this identity is always valid independent of the mobilities of the two bands. The second relates the Hall coefficient at B=0 to the density of the high mobility band, $R_0 = 1/(en_{high})$. This identity is mathematically correct in two independent cases: when the mobilities of the two bands are substantially different, or when the density of the second band is low enough $n_2 \ll n_1 \mu_1/\mu_2$, as is the case near the critical density. If our assumption of substantially different mobilities in the two bands is not accurate, this could modify our extracted densities much above the critical density, however, near the critical density the picture that we presented would still be completely accurate. Specifically, the determination of the universal density, which is a main point in this paper is completely robust and would not depend on any assumptions about the mobilities.

**A4.  Details of the calculation of the energy bands with atomic spin-orbit interactions**

Before introducing the spin-orbit interaction the three $t_{2g}$ orbitals of Ti are decoupled from each other and from the spin degree of freedom (ignoring the small term of second nearest-neighbor hopping). Therefore they are described by the diagonal Hamiltonian:

$$H_0 = \begin{bmatrix} \frac{k_x^2}{2m_h} + \frac{k_y^2}{2m_l} & 0 & 0 \\ 0 & \frac{k_x^2}{2m_l} + \frac{k_y^2}{2m_h} & 0 \\ 0 & 0 & \frac{k_x^2}{2m_l} + \frac{k_y^2}{2m_l} - \Delta_E \end{bmatrix}$$

where $m_l$ is the light mass, $m_h$ the heavy mass and $\Delta_E$ the splitting of the $d_{XY}$ from the other two due to the transverse confinement. The three dimensional Hilbert space represents the $d_{YZ}$, $d_{XZ}$, and $d_{XY}$ bands respectively. The same Hamiltonian in the full six dimensional Hilbert space, including spin, can be written as $\mathcal{H}_0 = H_0 \otimes \sigma^0$, where $\sigma^0$ is the 2 by 2 identity matrix. The atomic spin-orbit term is written in the same space as:

$$\mathcal{H}_{SO} = \Delta_{SO} L \cdot \sigma = \Delta_{SO} \sum_{\alpha=1}^{3} L^\alpha \otimes \sigma^\alpha,$$

where $\sigma^\alpha$ are the Pauli matrices acting on the spin and $L^\alpha$ are angular momentum-2 matrices, projected to the space of the three $t_{2g}$ orbitals. The band structure calculated from diagonalizing the total Hamiltonian $\mathcal{H}_0 + \mathcal{H}_{SO}$ is shown in Fig. S1. Because the spin-orbit coupling is completely local the Kramer's degeneracy at the atomic level is carried over to all wave-vectors.

Another type of coupling between the $t_{2g}$ orbitals can be induced because the transverse confinement is not symmetric to reflection $(z \rightarrow -z)$. This allows coupling between the $d_{XY}$ and $d_{XZ}$ through anti-symmetric hopping along y and similarly between the $d_{XY}$ and $d_{YZ}$ in hopping along x. Such couplings are encapsulated in the term:

$$\mathcal{H}_z = \Delta_z \begin{bmatrix} 0 & 0 & -i k_x \\ 0 & 0 & i k_y \\ i k_x & -i k_y & 0 \end{bmatrix} \otimes \sigma^0$$

Since this term is not fully local (i.e. it is momentum dependent), it splits the band degeneracy. In fact, taken together with the atomic-spin-orbit this last term generates an effective Rashba coupling. We note however that under realistic conditions $\Delta_z$ is much smaller than $\Delta_{SO}$ and therefore the Rashba splitting of the bands will be almost unnoticeable on the scale of Fig. S1.

### A5. Sample fabrication

As detailed in an earlier work[12], LaAlO$_3$ films were grown on TiO$_2$-terminated (001) SrTiO$_3$ single crystals by pulsed laser deposition at ~800 °C in ~$10^{-4}$ mbar of O$_2$. The repetition rate of the laser was 1Hz, with the fluence of each pulse being 0.6 J cm$^{-2}$. The film growth was monitored in situ using reflection high-energy electron diffraction. After growth, the samples were annealed in 200 mbar of O$_2$ at about 600 °C for one hour and cooled to room temperature in the same oxygen pressure. Hall bars were photolithographically patterned, and the sample was ultrasonically bonded using Al wire. For the high mobility sample, the substrate temperature was ~ 650 °C instead of 800 °C, and annealing took place at 530 °C as reported previously[1]. More information including the influence of oxygen pressure during growth can be found in ref. [13].